\documentclass[12pt,preprint]{aastex}
\input psfig.sty

\newcommand {\ea} {{\it et~al.}}
\newcommand {\be} {\begin{equation}}
\newcommand {\ee} {\end{equation}}

\def\MSUN{\rm M_{\odot}}
 
\def\MSUNYR{\rm M_{\odot}\,yr^{-1}}
\def\MDOT{\dot{M}}

\newbox\grsign \setbox\grsign=\hbox{$>$} \newdimen\grdimen \grdimen=\ht\grsign
\newbox\simlessbox \newbox\simgreatbox
\setbox\simgreatbox=\hbox{\raise.5ex\hbox{$>$}\llap
     {\lower.5ex\hbox{$\sim$}}}\ht1=\grdimen\dp1=0pt
\setbox\simlessbox=\hbox{\raise.5ex\hbox{$<$}\llap
     {\lower.5ex\hbox{$\sim$}}}\ht2=\grdimen\dp2=0pt
\def\simgreat{\mathrel{\copy\simgreatbox}}
\def\simless{\mathrel{\copy\simlessbox}}

\shorttitle{}

\shortauthors{Proga \ea}

\received{2004}
\begin{document}

\title{Dynamics of Line-Driven Disk Winds in Active Galactic Nuclei II:
Effects of Disk Radiation}

\author{Daniel~Proga\altaffilmark{1}$^,$\altaffilmark{2} and
Timothy~R.~Kallman\altaffilmark{3}$^,$\altaffilmark{4}}

\altaffiltext{1}{JILA, University of Colorado, Boulder, CO 80309-0440, USA}
\altaffiltext{2}{\tt{proga@colorado.edu}}
\altaffiltext{3}
{LHEA, GSFC, NASA, Code 662, Greenbelt, MD 20771, USA} 
\altaffiltext{4}{tim@xstar.gsfc.nasa.gov}

\begin{abstract}

 We explore consequences of a radiation driven disk wind model
 for mass outflows from active galactic nuclei (AGN).
 We performed axisymmetric time-dependent
 hydrodynamic calculations using the same computational technique as
 Proga, Stone and Kallman (2000).
 We test the robustness of radiation launching and acceleration
 of the wind for  relatively unfavorable conditions.
 In particular, we take into account the central engine
 radiation as a source of ionizing photons but neglect
 its contribution to the radiation force. Additionally,
 we account for the attenuation of the X-ray radiation by computing
 the X-ray optical depth in the radial direction assuming
 that only electron scattering contributes to the opacity.
 Our new simulations  confirm the main result from our
 previous work:
 the disk atmosphere can 'shield' itself  from external X-rays
 so that the local disk radiation can launch gas off the disk photosphere.
 We  also find  that the local disk force suffices to accelerate
 the disk wind to high velocities in the radial direction.  This is
 true provided the wind does not change significantly the geometry of 
 the disk radiation by  continuum scattering and absorption processes; 
 we discuss plausibility of this requirement.  Synthetic profiles of a typical
 resonance ultraviolet line  predicted by our models are consistent
 with observations of broad absorption line (BAL)  QSOs.

\end{abstract}


\section{INTRODUCTION}

Radiation pressure on spectral lines (line force) driving a wind from 
an accretion disk is the most promising hydrodynamical (HD) scenario 
for AGN outflows.  Within this framework, a wind is launched from the disk
by  the local disk radiation at radii where the
disk radiation is mostly in the ultraviolet (UV; e.g., 
Shlosman, Vitello \& Shaviv 1985; Murray et al. 1995, MCGV hereafter). 
Such a wind  is continuous and has mass loss rate and velocity 
which are are capable of explaining the blueshifted absorption 
lines observed from many AGN, if the ionization state is 
suitable. Such winds have the desirable feature that they do 
not rely on unobservable forces or fields for their motive power.  
However, detailed tests of this idea via modelling is challenging.  
The wind dynamics are coupled to the ionization and 
opacity properties of the gas, and the location and nature of the
radiation sources is not well understood.  Previous models by us
(Proga Stone and Kallman 2000, hereafter PSK) relied on 
UV radiation from the black hole in order to make the flow steady and 
to impart a strong radial component.  In this paper we report
new calculations which relax this requirement, and which demonstrate
quantitative consistency between disk winds and observations.

UV driven disk winds in AGN are motivated by analogy with winds from 
hot stars, which have been explored in great detail 
(eg. Lamers and Cassinelli 1999 and references therein). 
But they differ in important ways, including the role of rotation 
near a Keplerian disk, the non-uniform disk temperature 
distribution, and the influence of the strong X-ray flux from the 
inner disk and black hole.  Rotation acts to make the vertical 
component of gravity increase with height near the disk plane, and 
can also affect the wind trajectory when height is comparable to the 
radius.  Close to the disk plane the wind is driven by  photons emitted
locally, but at greater heights the radiation spectrum and the driving 
momentum depend on position; the increase in disk temperature at small 
radius can result in a net outward component to the momentum.
The wind density and mass loss rate can be estimated from the observed 
AGN UV luminosity.  When compared with the X-ray flux from the AGN, the  
density is low enough that the gas is predicted to  be highly ionized.  If so, 
the opacity needed for efficient driving and for line formation will 
be very small and the wind will fail.  On the other hand, estimates for the column 
density in the radial direction close to the disk surface 
are high enough that a portion of the wind can be shielded from this 
ionization.  Viable models for disk winds must account self-consistently 
for the wind driving, ionization, and self shielding.  MCGV pointed this out
and postulated the existence of `hitchhiking' gas which is not driven by 
UV but which provides the shielding.  Their calculation was a one-dimensional
time independent quasi-radial flow.  Proga, Stone \& Kallman (2000, PSK hereafter)
were able to to relax some of the MCGV assumptions, and 
explored consequences of a radiation driven disk wind model
by performing 2.5 dimensional  time-dependent HD simulations.
The most challenging component of fluid dynamics calculations
is treatment of the radiative transfer.  In the case of disk wind 
calculations this limitation forces an inexact treatment
of the radiative transfer, by assuming parameterized values for the 
X-ray and UV continuum opacities.

PSK found that a disk accreting onto a $10^8~\MSUN$ black hole 
at the rate of 1.1~$\rm g~s^{-1}$ ($1.8~\MSUNYR$) can launch a wind at $r_l\sim 10^{16}$~cm from 
the central engine. The X-rays from the central object are significantly 
attenuated by the disk atmosphere so they cannot prevent the UV 
radiation from pushing matter away from the disk. However, the X-rays can 
overionize the gas in the flow high above the disk and decrease the wind velocity.
For a reasonable X-ray opacity, e.g., $\kappa_{\rm X}=40~\rm g^{-1}~cm^2$, 
the disk wind can be accelerated by the central UV radiation to velocities of 
up to 15000~$\rm km~s^{-1}$ at a distance of $\sim 10^{17}$~cm from the central engine. 
The covering factor of the disk wind is $\sim 0.2$. The wind is unsteady
and  consists of an opaque, slow vertical flow near the disk that is 
bounded on the polar side by a high-velocity stream. A typical column 
density radially through the fast stream is a few $10^{23}~\rm cm^{-2}$ 
so that the stream is optically thin to the UV radiation but optically
thick to the X-rays.  

A key issue in modeling radiation pressure driven winds from disks 
is the relative importance of the driving by photons emanating from near 
the black hole, i.e. at radii smaller than the computational grid, 
compared with photons emitted by the disk at radii within the computational 
grid.  Models for CV disk winds, which resemble  AGN disk winds except for 
an absence of strong X-rays, show that strong steady flows require 
a central source of UV.  Motivated by this, PSK as well as MCGV 
considered the situation where the central radiation in the 
UV  accelerates the gas in the radial direction.
 A consequence of this assumption is that an efficiently driven 
wind must attenuate the X-rays, in order to avoid over-ionization, but must 
transmit the UV, in order to be driven.  This places an 
intrinsic limit on the radial column density, and hence on the mass loss rate, 
and can also influence the geometry of the flow.  

Our goal in the present paper is to  explore what happens when the assumption of 
a strong central flux is relaxed in the PSK models, and 
to examine  other consequences of the line-driven disk wind model. 
In particular, we check how robust is the disk wind solution and 
whether the solution predicts synthetic line profiles capable of 
reproducing line profiles observed in AGN.
In \S2, we summarize the key elements of our calculations.
We present the results from disk wind simulations and synthetic line
profile calculations in 
\S{2.1} and \S{2.2}, respectively. The paper ends in \S4, with our
conclusions and discussion.

\section{METHOD}

In this paper we extend work by PSK by relaxing some
of their assumptions and simplifications.
Our 2.5-dimensional HD numerical method is in most respects
as described by PSK. Here we only describe the key elements of the method and
list the changes we made. We refer a reader to PSK for details.

As in PSK, we apply line-driven stellar wind
models (Castor, Abbott \& Klein 1975, hereafter CAK) to  winds driven 
from accretion disks (see also Proga, Stone \& Drew 1998; 1999, PSD98 and
PSD99 hereafter). 
We specify the radiation field of the disk  by assuming that the temperature 
follows the radial profile of the optically thick accretion disk 
(Shakura \& Sunyaev 1973).
We  account for some of the effects of photoionization. In particular, 
we calculate the gas temperature assuming that the gas is optically thin 
to its own cooling radiation. 
We also take into account some of the effects 
of photoionization on the line force. In particular, 
we compute the parameters of the
line force using a value of the photoionization
parameter, $\xi$ and the analytical formulae due to Stevens \& Kallman
(1991). This procedure is computationally efficient and
gives approximate estimates for the number and opacity distribution of
spectral lines for a given $\xi$ without detail information about
the ionization state (see Stevens \& Kallman 1991).
Additionally, we take into account 
the attenuation of the X-ray radiation by computing the X-ray optical depth 
in the radial direction.  

We modify PSK's method as follows:
(i) we decrease the inner and outer radius of the computational domain
(see \S3 for details); 
(ii) to compute the radiation force due to lines, we use
the intensity of the radiation integrated over the UV band only 
(i.e., between $200~\AA$ and $3200~\AA$, see also Proga 2003).
(in PSK we took the UV flux to be a constant fraction of 
the flux from the black hole);
(iii) we 
exclude entirely the central object radiation force;
(iv) we compute the X-ray optical depth 
in the radial direction by considering only electron scattering
(in PSK we took the X-ray opacity to be $\kappa_X=40~\rm g^{-1}$ cm$^2$); 
(v) we do not explicitly treat continuum opacity for photons emitted from 
the disk, although the self-shielding of the lines is taken into account 
by our Sobolev treatment of the line force (in PSK we allowed the 
disk radiation to be attenuated 
in the radial direction with an opacity $\kappa_{UV}=0.4$~g$^{-1}$~cm$^2$).

The changes in our calculations extend the range of validity of 
calculations of PSK.  The decrease of the inner radius of 
the computation domain allows us to capture as much
as possible of the UV emitting disk within our grid.
The second change makes our calculations
as self-consistent as possible without
making them computationally prohibitive, i.e., 
there is no wind launching 
from very large radii where the disk is cold 
and radiates few UV photons, and from  very small radii where 
the disk effective temperature or gas temperature, or both, are too high  
to permit enough spectral lines.  The third change was motivated
by the results from PSK:  we want to test the importance of 
attenuation of the central UV flux as a  limit to the 
acceleration of  gas to high velocities, and also the 
extent to which the central flux is needed to make a strong and 
steady flow.  By neglecting the radiation force from the central object 
we are exploring relatively unfavorable conditions for wind acceleration.  
The only  force that can accelerate the wind is  the radiation force
due to the disk. 
The fourth change was also motivated
by our wish to explore relatively unfavorable conditions for
attenuation of X-rays by disk winds.  Using this treatment,  
our method gives a lower limit for the optical depth to ionizing 
photons.  The fifth change to the calculations means that
we allow all disk photons emitted toward a given point in the wind
to reach this point. This may correspond to an overestimate of the 
disk radiation force.  However, exact transfer of UV continuum 
is computationally prohibitive and is treated in the Sobolev approximation
even in one dimensional hot star wind  models.
We briefly discuss some of the consequences of our changes
in \S.4.

\section{RESULTS}

We present here results for the same model parameters as in PSK
with only two exceptions (see below). 
We assume the mass of the non-rotating black hole,  
$M_{BH}=10^8~\rm \MSUN$.
To determine the radiation field from the disk, we assume the mass accretion
rate $\MDOT_a=1.8$~M$_{\odot}$~yr$^{-1}$. 
These system parameters yield the disk luminosity, $L_D$ 
of 50\% of the Eddington luminosity and 
the disk inner radius, $r_\ast=3 r_S=8.8\times10^{13}$~cm, where 
$r_S=2GM_{BH}/c^2$ is the Schwarzschild radius of a black hole.

The radiation field from 
the central engine is specified by its luminosity $L_\ast=x L_D$
with $x$ set to 1, its fraction in the UV band
($f_{\rm UV}=0.9$) and its fraction in the X-rays ($f_{\rm X}=0.1$).
In PSK, $f_{\rm UV}=f_{\rm X}=0.5$
and these are the only changes to the model parameters we made.
The spectral energy distribution of the ionizing radiation is not
well known, our choice of values for $f_{\rm UV}$ and $f_{\rm X}$
is guided by the results from  Zheng et al. (1997) and Laor et al. (1997)
[e.g., see figure 6 in the latter, for a comparison
of the spectral energy distributions 
found for various samples of QSOs.]
As mentioned in \S2,  we include the central radiation only
as a source of ionizing photons and exclude  its contribution to
the radiation force.

There are two definitions of the photoionization parameter
used in literature: $\xi$ and $U$ (e.g., Krolik 1999). The former
is based on the ionizing flux while the latter on the number density
of the ionizing photons. For the adopted spectral energy distribution,
the conversion between the two is as follows: $\log U=\log \xi -1.75$.

Our computational domain is defined to occupy the radial range
$r_i~=~10~r_\ast \leq r \leq \ r_o~=~ 500~r_\ast$, 
and the angular range
$0^\circ \leq \theta \leq 90^\circ$. 
The $r-\theta$ domain is discretized into zones.  
Our  numerical resolution consists of 100 and 140
zones in the $r$ and $\theta$ directions, respectively. We use fixed zone size
ratios, $dr_{k+1}/dr_{k}=1.05$ and $d\theta_{l}/d\theta_{l+1} =1.066$.

\subsection{Two component disk wind solution}

Figure~1 shows the instantaneous density, temperature and 
photoionization parameter distributions and the poloidal velocity field
of the model.  The wind speed at the outer boundary is 
2000 to 12000~km~$\rm s^{-1}$.
This corresponds to a dynamical time of $\sim$0.2 yrs for the 
material at $\theta \simless 60^\circ$.  
Figure~1
shows results at the end of the simulation after $6.5$~years.  
Although the flow is still weakly time-dependent after this time has elapsed,
the gross properties of the flow (e.g., the mass loss rate and the radial
velocity at the outer boundary), settle down to steady time-averages.
As in the flow found by PSK, the wind has 3 components:
(i) a  hot, low density flow in the polar region
(ii) a dense, warm and fast equatorial  outflow from the disk, (iii)
a transitional zone in which the disk outflow is hot and struggles to
escape the system.  The main difference  with the results of PSK
is that here the transitional zone is much more prominent, and  it 
occupies  a large fraction of the computational domain.

In the polar region, the density is very small and close to the lower
limit that we set on the grid, i.e., 
$\rho_{min}=10^{-20}$~g~cm$^{-3}$.
The line force is negligible because the matter is highly ionized 
as indicated by a very large photoionization parameter ($\sim 10^8$). 
The gas temperature is close to the Compton temperature of the X-ray radiation.
The matter in the polar region is pulled by the gravity 
from the outer boundary, which is an artifact of the boundary conditions
(e.g., we do not model a jet which is likely to propagate 
through the polar region).
However, this region is sometimes filled with gas 
which is launched from the disk with a large vertical velocity 
in an episodic manner. 

The outflowing wind itself has distinct two components: hot and warm.
The two components are launched from the disk by the line force.
The gas density at the disk atmosphere and wind base is 
$\sim10^{-12}~{\rm g~cm^{-3}}$, so the photoionization parameter is low
(log($\xi$)$\leq$-5) despite the strong central 
radiation. However as the flow from the inner part of the disk is accelerated 
by the line force its density decreases. For the disk wind launched at
small radii ($r\simgreat r_i$),
this decrease of the density causes an increase in the gas temperature and 
the  photoionization parameter.  As this process proceeds  the gas becomes fully 
ionized and loses all driving lines.  The wind speed when this happens is 
not generally great enough to allow the gas to escape, and it tends to fall 
back toward the disk.  This `failed wind' has an effect on the remainder of the flow, 
both due to its shielding effect on the central X-rays and due to its pressure
as it falls toward the disk.

The disk wind launched at large radii does not become overionized downstream 
despite a density decrease. The photoionization parameter in this wind is 
kept low because of the large column density toward
the source of the ionizing photons. 
Consequently, the outer disk wind is not only launched but also accelerated 
by the disk line force.
We find that the outer wind becomes radial relatively close to the disk.
This may seem surprising because we set to zero the radiation force
in the radial direction due to the central object.
There are three effects responsible for the flow to be equatorial: 
(i)  overionization of 
the vertical part of the wind
 by the central object radiation;
(ii) the ram pressure of the failed wind launched at smaller
radii  that falls back after it is overionized; and 
(iii) the line force due to disk photons emitted 
interior to a given point in the wind. The two first effects dominate
at smaller radii and close to the interface between the hot
and warm flow whereas the third effect dominates 
above the disk at large radii where 
the fore-shortening of the disk radiation weakens.

The flows we find are time dependent and have 
density and velocity fluctuations of $\sim 10^2$ even after the flow has 
evolved for many dynamical times. In particular, there are occasionally
regions along the equator with the gas densities 
lower than those for hydrostatic equilibrium. From those low density regions,
the radiation force can launch a wind with an acceleration
length scale shorter than for a CAK-like wind [the line force
is a strong function of the gas density (CAK)]. In fact, we observe
that the line acceleration
can be so efficient that some gas can reach velocities comparable 
to the escape velocity before it is eventually overionized.
This flow behavior manifests itself as erratic 
high velocity ejections of gas from the inner disk.

Figure~2 presents the run of the density, radial velocity, mass flux density,
accumulated mass loss rate, photoionization
parameter and column density as a function of  the polar angle, $\theta$, at
the outer boundary, $r=r_o=4.422 \times 10^{16}$~cm from Figure~1.
The accumulated mass loss rate, $\dot{m}$ and the column density, $N_H$ 
are computed as in PSK (see eq. 13 in PSD~98 and eq. 24 in PSK for 
the definitions of $\dot{m}$ and $N_H$, respectively).
The gas density is very low, i.e., 
$\rho_{min}=10^{-20}~{\rm g~cm^{-3}}$,
for $\theta$ between $0^\circ$ and $5^\circ$. Then 
the gas density increases  with $\theta$ between $5^\circ$ and $45^\circ$
to the level of a few $\times 10^{-18}~{\rm g~cm^{-3}}$.
For  $45^\circ \simless \theta \simless 55^\circ$, the density decreases
to the level of $\rho_{min}=10^{-20}~{\rm g~cm^{-3}}$ at $\theta=55^\circ$.
Then the density increases again with $\theta$ and reaches the maximum
of  a few $\times 10^{-18}~{\rm g~cm^{-3}}$ at $\theta=75^\circ$.
This is followed by a decrease of the density with the minimum of 
a few $\times 10^{-20}~{\rm g~cm^{-3}}$ at $\theta=88^\circ$.
For, $\theta > 88^\circ$, the density 
sharply increases, as  might be expected
of a density profile determined by hydrostatic equilibrium.
The radial velocity is $-2000 $~km~$\rm s^{-1}$ 
for $0^\circ  \simless \theta \simless 5^\circ$
and has a broad flat maximum at the level of $3000$~km~$\rm s^{-1}$
for $\theta$ between $5^\circ$ and $55^\circ$.
For $\theta$ between $55^\circ$ and $60^\circ$, the radial velocity
is negative 
and then gradually increases to $20000 $~km~$\rm s^{-1}$ at $\theta \simless
85^\circ$. This is followed by  a drop of $v_r$ to a very small value
near the equator.
We note that  $v_r$
stays nearly constant at the level of   $12000 $~km~$\rm s^{-1}$
for $72^\circ \simless \theta \simless 82^\circ$.

The accumulated mass loss rate is negligible for $\theta \simless 35^\circ$
because of the very low gas density and velocity.
For $\theta \simgreat 35^\circ$, $d\dot{m}$ increases 
to $\sim 1\times 5\times10^{23}~\rm g~s^{-1}$ 
at $\theta \approx 45^\circ$.  For $\theta$ between $45^\circ$ and
$67^\circ$,  the accumulated mass loss rate stays nearly constant.
For $\theta \simgreat 77^\circ$, the accumulated mass loss rate
increases to $\sim 1.4\times 10^{25}~\rm g~s^{-1}$ at $\theta\approx 85^\circ$.

The column density in the wind increases gradually with $\theta$.
In particular, it increases from $\sim 10^{20}~\rm cm^{-2}$ at $\theta =
18^\circ$ to $10^{24}~\rm cm^{-2}$ at $\theta = 60^\circ$. 
For $\theta > 60^\circ$, $N_H$  continues to increase and indicates that
the wind is optically thick to electron scattering at $r_o$
for the central object radiation.
The column densities greater than $10^{25}~\rm cm^{-2}$
are effectively infinite, and represent complete obscuration of the central
object. Winds with such high $N_H$ can still be radiation driven owing 
to the difference between the column the disk radiation sees and the column
the central source radiation sees.
We expect that the disk wind obscuration
can change the ratio
between the number of all QSOs and BAL QSOs, i.e., QSOs, viewed by an observer
almost edge-on, would not be detected or identified as QSO.

The photoionization parameter is very high, $\sim 10^8$, for 
$\theta \simless 60^\circ$ because of the very low density in the polar
region. However,
over next $7^\circ$, $\xi$ 
drops by many order of
magnitude owing to the increase of the column density.
The $\theta$ profiles of the flow properties show that the overoinized
outflow can contribute to the total mass loss rate at the level
of a few per cent.

Our new simulations of AGN disk wind confirm the main result
from PSK:
the disk atmosphere can 'shield' itself 
so that the local disk radiation can launch gas off the disk
photosphere. Here we find that this results holds even
when the condition for shielding are unfavorable: the ionizing
photons are only scattered on electrons but not absorbed by the shielding
gas. 
Our new simulations also provide some new insights to
the acceleration. In particular, we find 
the local disk force suffices to accelerate
the disk wind to high velocities in the radial direction provided
the wind does not change the geometry of the disk radiation by 
continuum scattering and absorption processes.

The main difference between our disk wind solution
and the solution presented in PSK is the properties and behavior of 
the hot component of the disk outflow which struggles to escape the system.
In PSK, this component occupied a relatively narrow $\theta$ range
($\Delta \theta \approx 5^\circ$) above the disk wind, whereas
here $\Delta \theta \approx 40^\circ$.
The key reason for this difference is the strength of the radial line
force relative to the latitudinal line force acting on the inner disk wind.
In PSK, the radial force is strong and accelerates the inner wind
on a relatively long length scale (of order of the wind
launch radius).  A  small fraction
of the inner flow is overionized and fails to develop into a
strong outflow. This overionized gas falls on the freshly
launched disk wind and introduces weak perturbations to the disk
wind so that the wind is not significantly slowed down nor disrupted.
Here,  on the other hand, the radial force
is relatively weak and the flow from the innermost disk tends to be 
vertical. 
The flow geometry is important for how long the 
flow can stay shielded: for the vertical geometry, the flow height
can be larger than the height of the shielding
gas whereas for the radial geometry, the flow height
can stay smaller than the height of the shielding
gas even at very large distances from the launch point.
Therefore vertical flow is more susceptible to overionization 
than radial flow;
a much larger fraction of the inner flow is 
overionized in the vertical flow than in the radial flow. In 
our simulations the overionized flow
is relatively dense and as it falls on the disk  it disrupts
the freshly launched the disk wind.  

We note that the mass loss rate in our simulations is
smaller than in PSK's simulation by a factor of $\sim$ 2.
This is an unexpected result because here we allow
launching a wind from radii smaller than those in PSK and 
a simple scaling law
for line-driven disk wind models suggests that mass
loss rate should increase with decreasing radius (e.g., PSD98;
Proga \& Kallman 2002). Additionally, one can
argue that because of this scaling, overionization
should become less of a problem with decreasing radius (Proga \& Kallman
2002). What is then responsible for this low mass loss rate?
Our analysis of the flow properties and time evolution 
shows that at the early phase of the evolution, the mass flux
density is consistent with the expectations based on the above arguments. 
However, as the inner disk wind becomes overionized and starts falling back
toward the equator the mass loss rate decreases because the average 
gas density at the wind base increases which reduces the driving line force. 
Additionally, the infalling
gas is dense and can significantly slow the outflow
which it encounters on its way toward the disk. There are then
a few factors which reduce the mass loss rate but
the primary factor is the flow overionization.

Our time-dependent disk wind solutions
differ from those found by PSD98, in which it was shown that
CV disk winds are unsteady in time when the wind is driven solely by 
disk radiation.
In the latter, the flow could also fall back on the disk, but  
it was not strong enough to cause a significant decrease in 
the mass loss rate.  Furthermore, in the PSD98 
simulations there was no radiation
capable of fully ionizing the wind so that regions of relatively low
density, above the falling gas,
could be accelerated to high velocities
and contribute to the total mass loss rate. 

\subsection{UV absorption line profiles}

Simulations presented in PSK and here can serve as a proof-of-concept
for the radiation driven disk wind model of outflows in AGN. 
Some consequences
of the radiation driven disk wind model  have been explored 
in the context of CV and low mass X-ray binaries 
(LMXB; e.g., see PSD98 and Proga \& Kallman 2002, respectively).
Generally, the insights from those models allow us to explain 
systems with strong evidence for UV absorbing disk wind 
-- such as CVs -- and systems without 
UV absorbing disk wind -- such as LMXB. In the case of CVs, the model also
predicts UV line profiles which are capable of reproducing observations
(Proga 2003). In this section, we present synthetic line profiles
predict by our model and briefly discuss
their relevance to AGN observations.

We present line profiles for the C~IV~$\lambda 1549$ transition.
We compute the profiles by integrating flux from the entire
disk at a given orientation (i.e., as in the dynamical models,
the continuum source has a finite size).
The line profile calculation performed here
are exactly as in Proga et al. (2002)
with the exception that we set the C~IV abundance to one in the regions with 
the gas temperature from 8000~K to 420000~K and to zero elsewhere.
For the adopted model parameters, the upper limit 
for the gas density corresponds to the disk
effective temperature at $r_i$.
We refer a reader to Proga et al. (2002) for details 
on line-profile calculations. 

The C~IV abundance could be in principle, computed 
self-consistently for given wind structure and radiation field. However,
we save this for a future paper and focus on gaining some insights
from the multidimensional models of  disk winds and predicted profiles. 

To illustrate the effect of viewing angle on the line profiles,
Figure~3 shows line profiles for five
inclination angles: $i=~60^\circ, 65^\circ, 70^\circ, 75^\circ$, and 
$80^\circ$. 
We limit our line profile calculation to the absorption contribution
only. Modeling of the emission  component depends on the influence of 
thermal line emission, and we will discuss this in a future paper.
Figure~3 clearly shows that the line profiles are very sensitive
to the inclination angle. In particular, for $i=65^\circ$
the line absorption is broad and nearly symmetric whereas
for $i$ just $5^\circ$ higher a strong blueshifted absorption
dominates the line profile (compare figs 3c and 3d). 
We note that for $i=60^\circ$
the profiles is more symmetric than for $i=65^\circ$.
In the latter case, the blueshifted absorption is somewhat stronger
than the redshifted absorption.
Overall our line profiles are representative of
generic line profiles for a bipolar wind from a rotating disk (e.g.
Proga et al. 2002 and references therein).
For example, for a relatively high inclination angles ($i\simgreat
50^\circ$) the profiles are affected by the absorption
in the rotating base of the wind, so  blueshifted as well 
as redshifted absorption is present (see Proga et
al. 2002, and references therein). The presence of the expanding
wind manifests itself in the profiles by a blueshifted absorption 
for inclination angles where the continuum source is viewed through
the expanding wind. In our wind model, the disk 
wind with a low $\xi$ (i.e., high C~IV abundance) has a half-opening angle of $\sim 65^\circ$.
Therefore only for $i\simgreat 60^\circ$, our line profiles 
show blueshifted absorption stronger than redshifted absorption.

In general, the line profiles predicted by our models are similar to
those predicted by the model for CV winds (Proga 2003).
However, there are two important differences: (i)
for low $i$, CV wind models predict a blueshifted absorption
whereas here little absorption is present at those angles and (ii)
for all $i$, CV wind models predict that the maximum of the blueshifted
absorption is line-centered whereas here the maximum 
can be blueshifted by as much as $5000$~km~$\rm s^{-1}$.
The reason for these differences is the fact that in CV wind models,
the base of the wind covers the whole
continuum source whereas here the continuum source extends inside
the computational domain and therefore the continuum is only partially 
covered by the flow.

Our line profiles are consistent with many of the properties of  profiles
observed in broad absorption line QSOs (e.g., Korista et al. 1993; Hall et
al 2002).
In particular, the synthetic line profiles can be as strong and broad
(up to 30000 ~km~$\rm s^{-1}$) as those observed. Additionally, for certain 
inclination angles the maximum of the blueshifted absorption
is far from the line center - this is consistent with so-called
detached troughs observed in some QSOs. Some of the profiles
shown in Fig. 3 display  prominent redshifted absorption. 
In observed lines, this component is likely to be affected 
by the emission contribution to the line (e.g., Proga 2003). 
The line emission can fill in the redshifted
absorption and also some of the highly blueshifted absorption 
produced by the wind near the equator where the rotational
velocity exceeds the expansion velocity. For example, 
Proga (2003) showed that
for high inclinations ($i> 80^\circ$), the profile has
double-peaked emission (figs. 1e and 1j in Proga 2003). 

\section{CONCLUSIONS AND DISCUSSION}

In this paper we have presented numerical simulations of the
radiation-driven disk wind scenario for AGN outflows.  These 
models provide insights into the wind's capability to shield
itself from ionizing radiation. 
We performed axisymmetric time-dependent 
hydrodynamic calculations similar to those of Proga, Stone and Kallman 
(2000).  We studied the robustness of the radiation launching and 
acceleration of the wind for  relatively unfavorable conditions.
In particular, we have taken into account the central engine
radiation as a source of ionizing photons but neglected
its contribution to the radiation force. Additionally,
we have accounted for the attenuation of the X-ray radiation by computing 
the X-ray optical depth in the radial direction assuming
that only electron scattering contributes to the opacity. 
Our new simulations  confirm the main result
from PSK: the disk atmosphere can 'shield' itself  from external X-rays
so that the local disk radiation can launch gas off the disk
photosphere.  We have also found  that the local disk force suffices to accelerate
the disk wind to high velocities in the radial direction.  This is true provided
the wind does not change significantly the geometry of the disk radiation by 
continuum scattering and absorption processes, and we discuss the
plausibility of this requirement. Overall our models are consistent
with observations of broad absorption line (BAL)  QSOs. Additionally,
synthetic profiles of a typical resonance ultraviolet (UV) line, predicted
by our models, promise to reproduce observed BALs and it is likely
that the model can account for other AGN outflows observed 
in UV as well as in X-rays.

Strong ionizing radiation has long been recognized as a serious
challenge to theories of AGN outflows absorbing the UV radiation.
Radiation-driven disk wind models are particularly affected by this problem
because the overionization of the wind can prevent the wind from
being driven. But in other models, including a magnetic one,
overionization also needs to be addressed owing to the requirement for 
sufficient un-ionized gas to account for the observed lines.
Here we find that an indirect effect of overionization
can reduce but not prevent the development of a wind,
for our choice of  model parameters: the so-called failed inner
disk wind. If the density of the failed wind is relatively
high then the ram pressure of such a  wind can seriously slow
down the disk wind which otherwise could be driven by radiation.
We emphasize that this damaging effect of the failed wind
can be important to explain AGN winds within the radiation
driven disk wind scenario rather than being used
as an argument against the scenario. In particular,
in this scenario the accretion disk, with parameters as adopted
here (see \S3), should produce a powerful wind from radii as small
as $10^{14}-10^{15}~{\rm cm}$ because the radiation at those
radii is mostly in the UV. However, such a wind 
would be too fast to be consistent with AGN observations.
One of plausible solutions of this problem is to allow
for the energy dissipation  at small radii to occur not 
only inside but also outside the disk 
(i.e., in the disk corona). Consequently, 
the UV radiation
from the innermost disk would be reduced while ionizing radiation would be increased.
The two factors would reduce launching of the wind. A complimentary
solution is that the failed wind 
can disrupt development of a wind up to a radius of a few 
$\times 10^{15}~{\rm cm}$.

We have performed many simulations to study the sensitivity 
of disk wind solutions to the black hole mass and disk luminosity.
These results will appear in Proga et al.(2004) (in
preparation). Here we only mention that the wind solution is
very sensitive to the luminosity compared to the Eddington
limit. In particular, we find that for the model parameters
as in \S.3 but with
the luminosity reduced by a factor of 5 (from 50\% to 10\% of the Eddington
limit) there is no disk wind.
The primary reason for this luminosity sensitivity is 
the fact that the mass flux density of the wind
decreases strongly with decreasing disk luminosity 
and the wind is more subject to overionization (e.g. Proga \& Kallman 2002).
Our simulations show also that for a fixed Eddington fraction
(e.g., 50\%) it is easier to produce a wind for 
$M_{\rm BH} \simgreat 10^7~\MSUN$ than for 
$M_{\rm BH} \simless 10^7~\MSUN$. This result is a consequence of
the decrease of the UV contribution to the  disk total radiation with
decreasing mass of the BH in the Shakura Sunyaev disk model
for a fixed  Eddington fraction.

We have also performed many simulations to study the sensitivity of 
our main results 
to the model parameters. The most important parameter of the model
is the X-ray opacity. This quantity could be computed from first principles
for given spectral energy distribution, strength
and geometry of the radiation, and the structure and chemical composition
of the flow.
However, such an approach to $\kappa_{\rm X}$'s calculations 
is not computationally feasible 
and therefore we have explored asymptotic approximations, 
e.g., assuming that $\kappa_{\rm X}=0.4~\rm g^{-1}~cm^2$ 
(corresponding to the Thomson cross section).
As we discussed above, such a simplistic approach yields rather conservative
results for the wind strength. In our sensitivity
studies, we have computed wind models using a more realistic
$\kappa_{\rm X}$. In particular, 
we computed $\kappa_{\rm x}$ based on the analytic fit
to the detailed photoionization calculations using the photoionization code 
XSTAR.
We find that $\kappa_{\rm x}$ can be expressed a function of 
the photoionization parameter, $\xi$ in the following way: 
$\kappa_{\rm X}=\min(10^3, 4\times10^3 \xi^{-1.2}+1)$, in units of 
$0.4~{\rm g^{-1} cm ^2}$. We found that as expected a wind
with this higher $\kappa_{\rm x}$ is stronger (e.g., in terms
of $\MDOT$  and the opening angle) the model solution described here.

The radiative transfer of continuum photons inside 
the disk wind is as critical
as the  wind photoionization structure in modeling disk winds. 
To accelerate the wind to velocities comparable to the escape velocity 
we should not rely on the radiation which originates
interior to the wind zone because
the wind will shield itself from this radiation in the same
way as it shields itself from the external ionizing radiation.
Here we explore a  configuration where the ionizing
radiation comes from a point source located at the center
while the driving radiation comes from the extended source,
(i.e., the UV emitting disk). Calculations of the radiative
transfer of the former is relatively simple as it can be approximated
by 1D solutions for the optical depth. However, the radiative
transfer of the latter is intrinsically 3D and computationally expensive.
We argue that the UV continuum photons emitted from the disk
into the overlying disk wind can propagate through the disk
without large changes of the net radiation flux, hence our assumption
that the wind is optically thin to the disk photons. On the other hand,
X-rays which are emitted outside the disk wind  suffer many scatterings
and absorptions once they reach the wind zone. Consequently, the
X-rays do not penetrate deep inside the wind zone. The X-ray transfer
in the case we consider is analogous to the problem of disk irradiation 
by an external source whereas the UV transfer is analogous
to the problem of radiation transfer in an extended atmosphere.

Our simulations show how the line force can produce a fast, 
highly ionized wind: if the density near the wind base is relatively low
and the base is shielded over long enough length scale above the disk
then the local disk radiation can accelerate gas to relatively high velocities
(even exceeding the escape velocity) before the gas becomes overionizated.  
At large radii, such a wind would be transparent to the UV radiation but may
be able to absorb X-rays. 
We plan to present models for the X-ray properties of disk winds in a future 
paper.

We thank N. Arav, M.C., Begelman, J.R. Gabel, P.B. Hall, G.T. Richards, 
and J.M. Stone for useful discussions. 
We also thank  an anonymous referee for comments
that helped us clarify our presentation.
We acknowledge support provided by NASA through grant 
HST-AR-09947 from the Space Telescope Science Institute, which is operated 
by the Association of Universities for Research in Astronomy, Inc., 
under NASA contract NAS5-26555. DP also acknowledges support from 
NASA LTSA grants NAG5-11736 and NAG5-12867.
We acknowledge support from the W. M. Keck Foundation, 
which purchased the JILA 74-processor
Keck Cluster.

\eject

\begin{figure}
\begin{picture}(600,500)
\put(0,0){\includegraphics{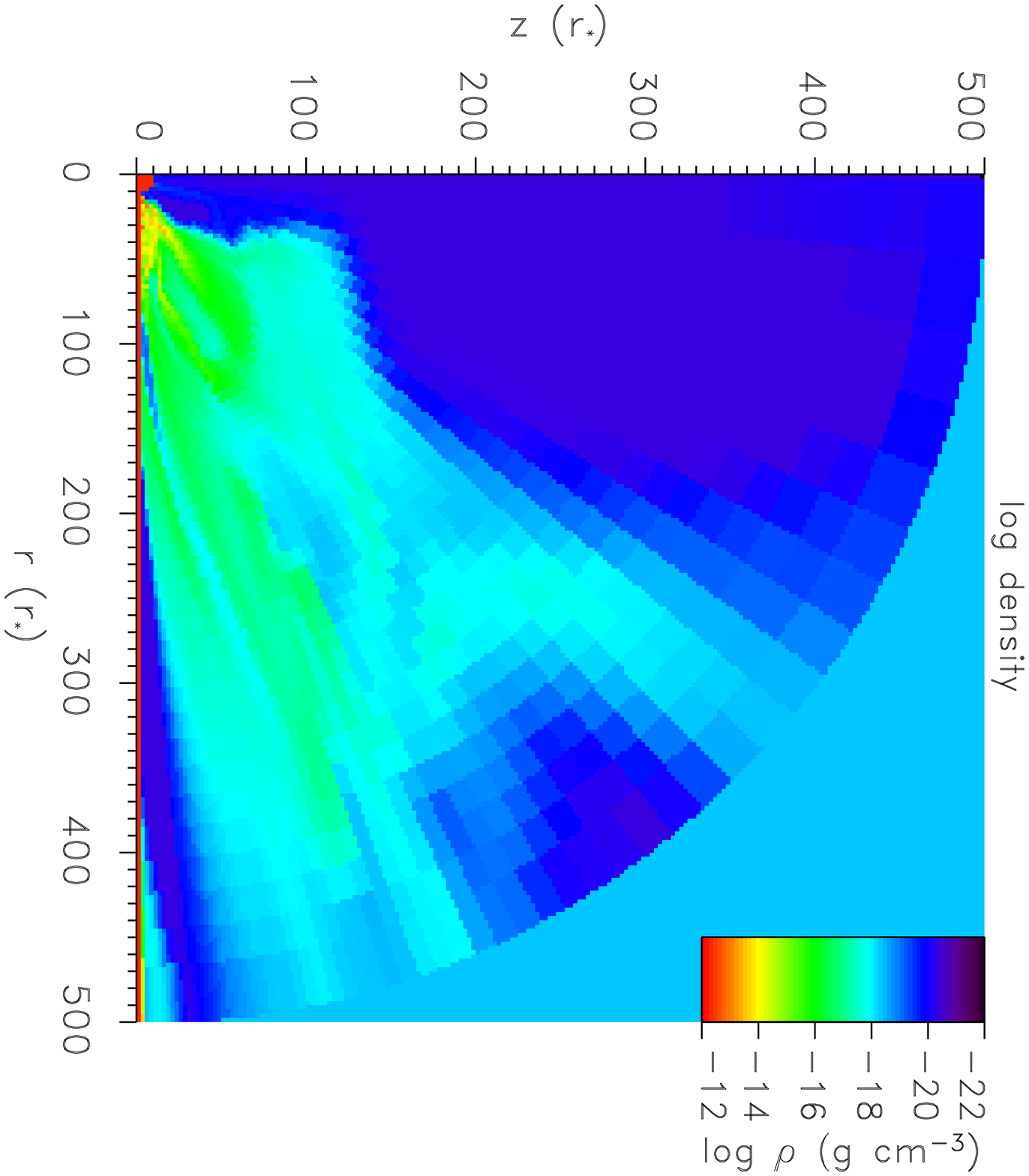}}

\put(0,0){\includegraphics{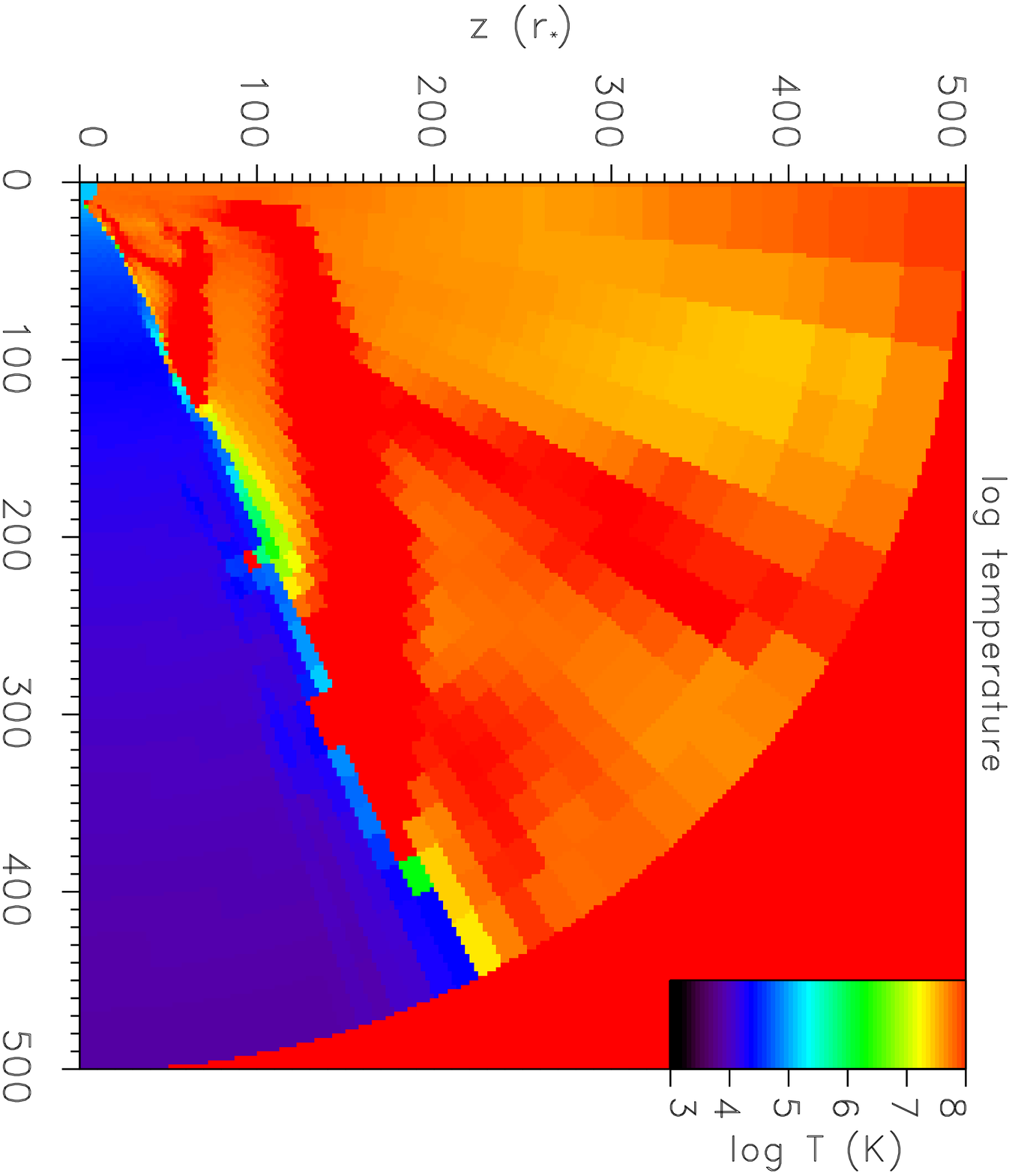}}

\put(0,0){\includegraphics{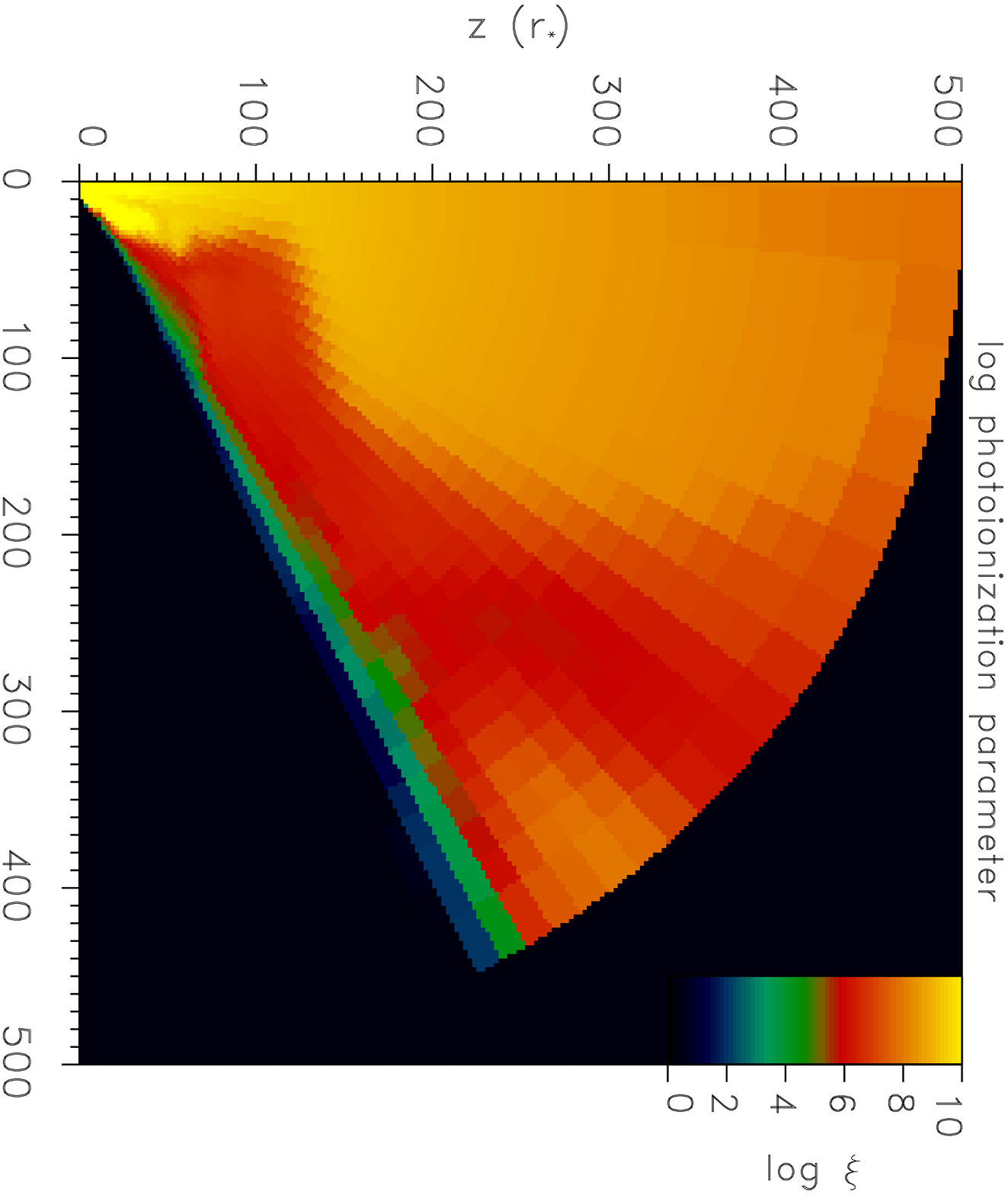}}

\put(0,0){\includegraphics{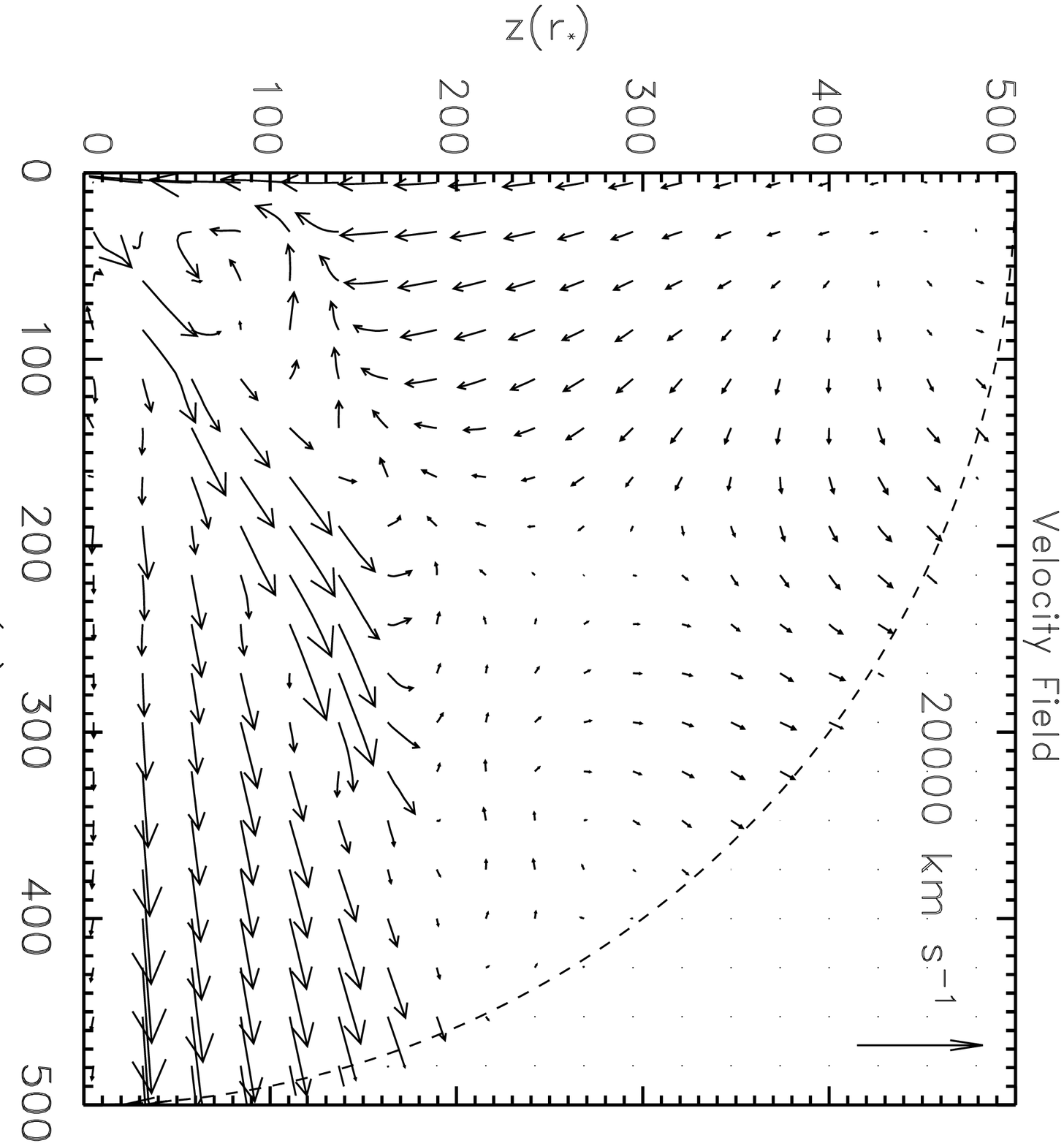}}
\end{picture}
\caption{
The top left panel is a color density map of the AGN disk wind
model, described in the text. The top right panel is a color
gas temperature map of the model while the bottom left panel
is a color photoionization parameter map. Finally, the bottom right
panel is a map of the velocity field (the poloidal component only). 
In all panels the rotation axis of the disk is along the left hand vertical 
frame, while the midplane of the disk is along the lower horizontal frame.
}
\end{figure}

\begin{figure}
\begin{picture}(600,500)
\put(0,0){\includegraphics{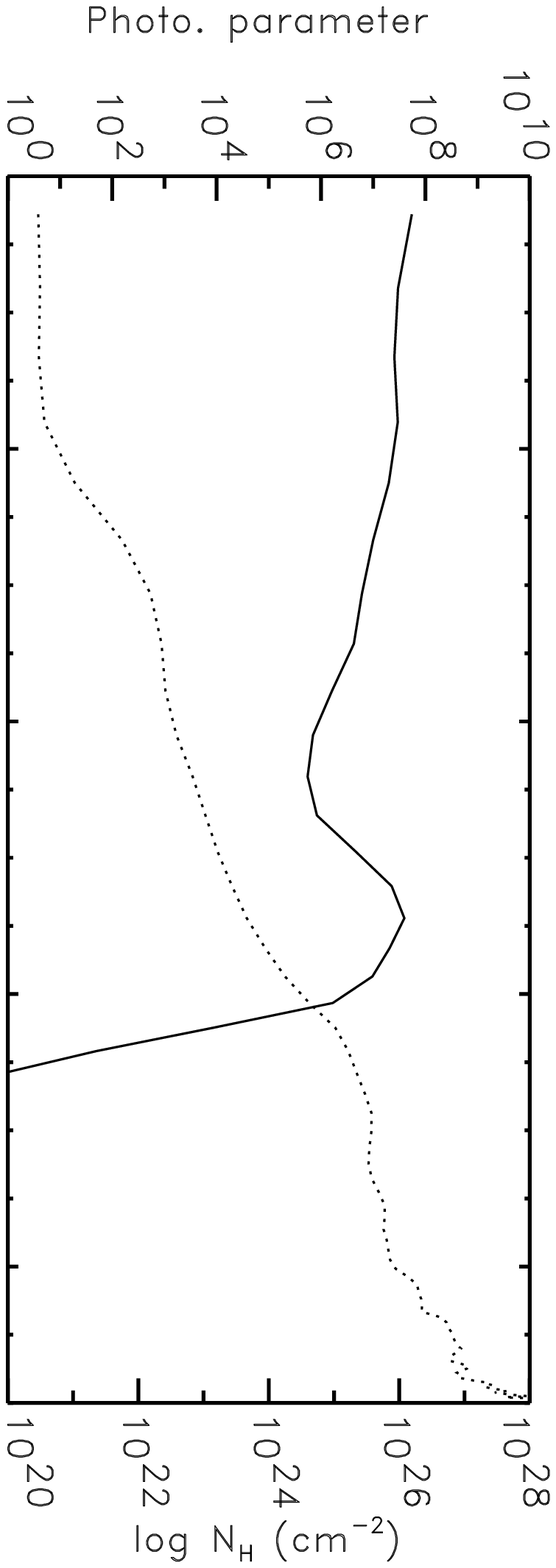}}

\put(0,0){\includegraphics{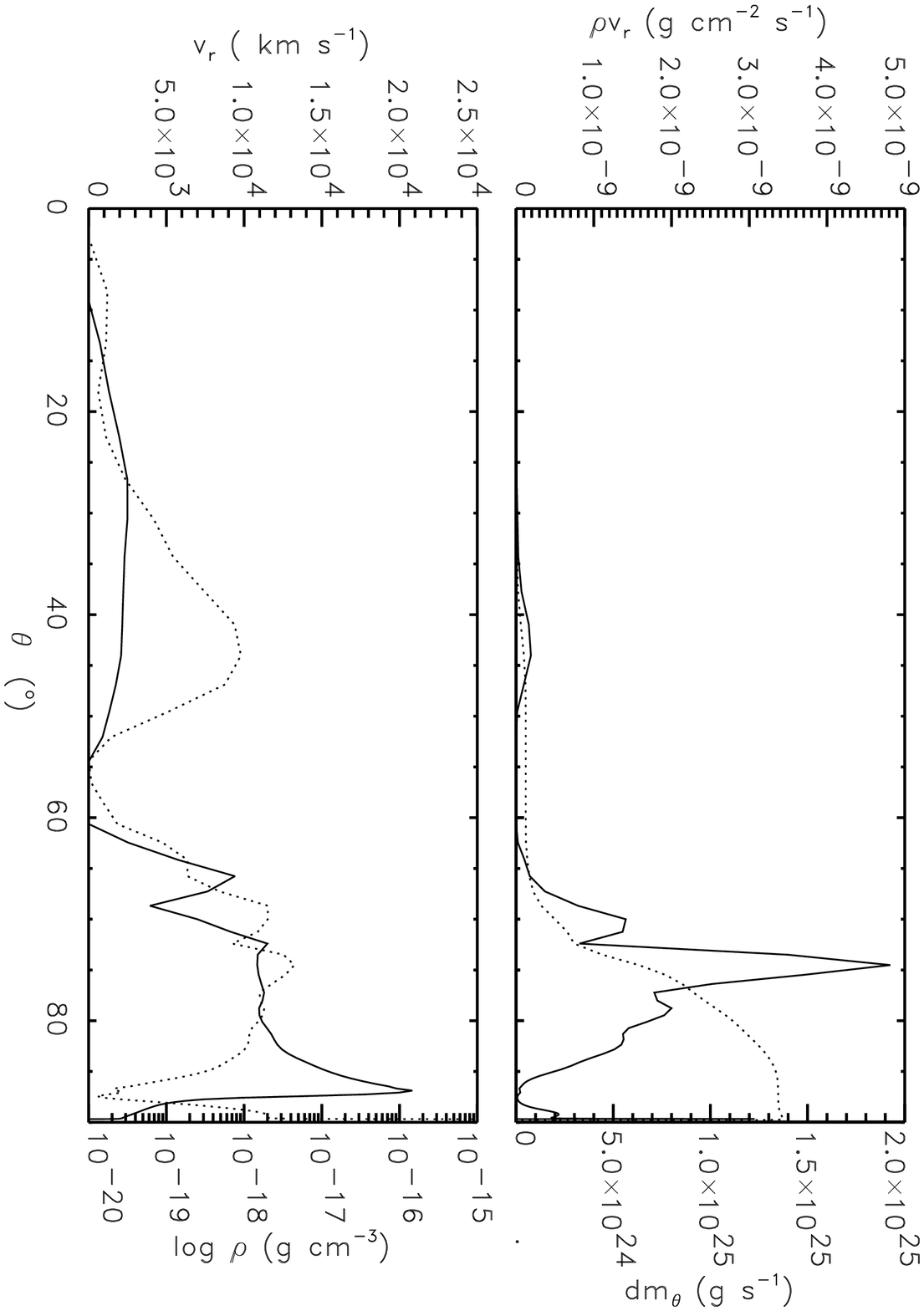}}
\end{picture}
\caption{Quantities at the outer boundary in our model.
The ordinate on the left hand side of each panel refers to the solid
line, while the ordinate on the right hand side refers to the dotted
line. 
The column density, $\rm N_H$ is calculated  along the radial direction.}
\end{figure}

\begin{figure}
\begin{picture}(180,490)

\put(0,0){\includegraphics{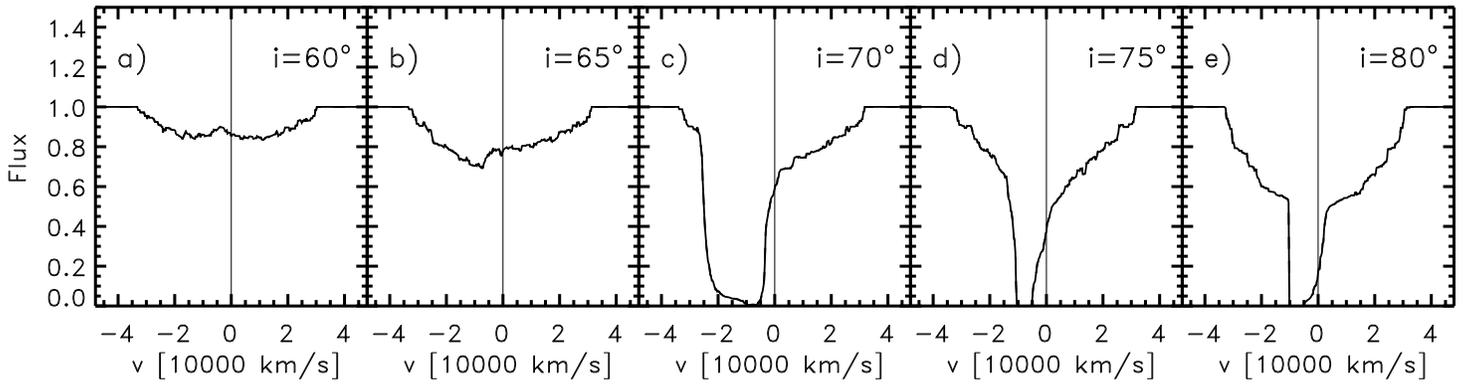}}

\end{picture}
\caption{Profiles of a resonance ultraviolet line  
for our hydrodynamical disk wind models a a function
of inclination angle, $i$ (see top right corner of each panel 
for the value of $i$).
The zero velocity corresponding to the line center is indicated by the vertical
line.}
\end{figure}

\end{document}